\documentclass[twoside,twocolumn,9pt]{article}
\usepackage{extsizes}
\usepackage[super,sort&compress,comma]{natbib} 
\usepackage[version=3]{mhchem}
\usepackage[left=1.5cm, right=1.5cm, top=1.785cm, bottom=2.0cm]{geometry}
\usepackage{balance}
\usepackage{mathptmx}
\usepackage{sectsty}
\usepackage{graphicx} 
\usepackage{lastpage}
\usepackage[format=plain,justification=justified,singlelinecheck=false,font={stretch=1.125,small,sf},labelfont=bf,labelsep=space]{caption}
\usepackage{float}
\usepackage{fancyhdr}
\usepackage{fnpos}
\usepackage[english]{babel}
\addto{\captionsenglish}{%
  
}
\usepackage{array}
\usepackage{droidsans}
\usepackage{charter}
\usepackage[T1]{fontenc}
\usepackage[usenames,dvipsnames]{xcolor}
\usepackage{setspace}
\usepackage[compact]{titlesec}
\usepackage{hyperref}
\usepackage{braket}
\usepackage{bm}
\usepackage{tikz}
\usepackage{tikzscale}
\usetikzlibrary{positioning, calc}

\usepackage{epstopdf}%This line makes .eps figures into .pdf - please comment out if not required.

\definecolor{cream}{RGB}{222,217,201}

\begin{document}

\pagestyle{fancy}
\thispagestyle{plain}
\fancypagestyle{plain}{
%%%HEADER%%%
\renewcommand{\headrulewidth}{0pt}
}
%%%END OF HEADER%%%

%%%PAGE SETUP - Please do not change any commands within this section%%%
\makeFNbottom
\makeatletter
\renewcommand\LARGE{\@setfontsize\LARGE{15pt}{17}}
\renewcommand\Large{\@setfontsize\Large{12pt}{14}}
\renewcommand\large{\@setfontsize\large{10pt}{12}}
\renewcommand\footnotesize{\@setfontsize\footnotesize{7pt}{10}}
\makeatother

\renewcommand{\thefootnote}{\fnsymbol{footnote}}
\renewcommand\footnoterule{\vspace*{1pt}% 
\color{cream}\hrule width 3.5in height 0.4pt \color{black}\vspace*{5pt}} 
\setcounter{secnumdepth}{5}

\makeatletter 
\renewcommand\@biblabel[1]{#1}            
\renewcommand\@makefntext[1]% 
{\noindent\makebox[0pt][r]{\@thefnmark\,}#1}
\makeatother 
\renewcommand{\figurename}{\small{Fig.}~}
\sectionfont{\sffamily\Large}
\subsectionfont{\normalsize}
\subsubsectionfont{\bf}
\setstretch{1.125} %In particular, please do not alter this line.
\setlength{\skip\footins}{0.8cm}
\setlength{\footnotesep}{0.25cm}
\setlength{\jot}{10pt}
\titlespacing*{\section}{0pt}{4pt}{4pt}
\titlespacing*{\subsection}{0pt}{15pt}{1pt}
%%%END OF PAGE SETUP%%%

%%%FOOTER%%%
\fancyfoot{}
\fancyfoot[LO,RE]{\vspace{-7.1pt}\includegraphics[height=9pt]{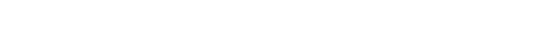}}
\fancyfoot[CO]{\vspace{-7.1pt}\hspace{11.9cm}\includegraphics{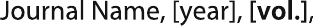}}
\fancyfoot[CE]{\vspace{-7.2pt}\hspace{-13.2cm}\includegraphics{head_foot/RF}}
\fancyfoot[RO]{\footnotesize{\sffamily{1--\pageref{LastPage} ~\textbar  \hspace{2pt}\thepage}}}
\fancyfoot[LE]{\footnotesize{\sffamily{\thepage~\textbar\hspace{4.65cm} 1--\pageref{LastPage}}}}
\fancyhead{}
\renewcommand{\headrulewidth}{0pt} 
\renewcommand{\footrulewidth}{0pt}
\setlength{\arrayrulewidth}{1pt}
\setlength{\columnsep}{6.5mm}
\setlength\bibsep{1pt}
%%%END OF FOOTER%%%

%%%FIGURE SETUP - please do not change any commands within this section%%%
\makeatletter 
\newlength{\figrulesep} 
\setlength{\figrulesep}{0.5\textfloatsep} 

\newcommand{\topfigrule}{\vspace*{-1pt}% 
\noindent{\color{cream}\rule[-\figrulesep]{\columnwidth}{1.5pt}} }

\newcommand{\botfigrule}{\vspace*{-2pt}% 
\noindent{\color{cream}\rule[\figrulesep]{\columnwidth}{1.5pt}} }

\newcommand{\dblfigrule}{\vspace*{-1pt}% 
\noindent{\color{cream}\rule[-\figrulesep]{\textwidth}{1.5pt}} }

\makeatother
%%%END OF FIGURE SETUP%%%

%%%TITLE, AUTHORS AND ABSTRACT%%%
\twocolumn[
  \begin{@twocolumnfalse}
{\includegraphics[height=30pt]{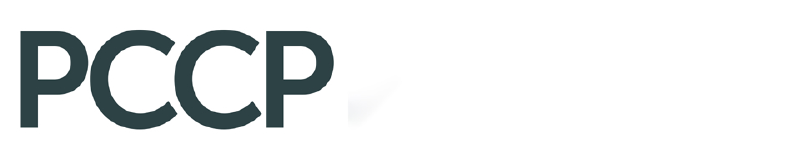}\hfill\raisebox{0pt}[0pt][0pt]{\includegraphics[height=55pt]{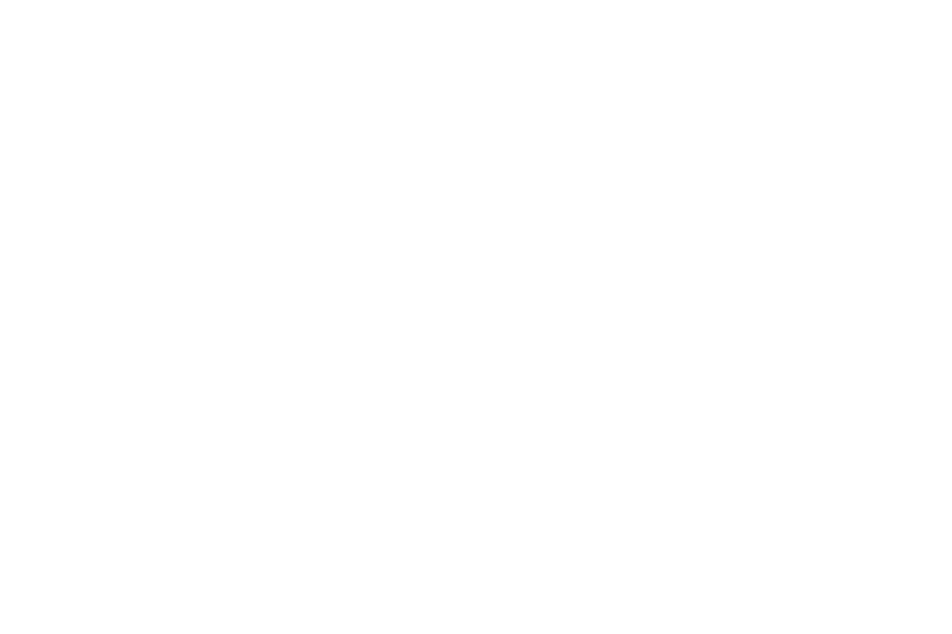}}\\[1ex]
\includegraphics[width=18.5cm]{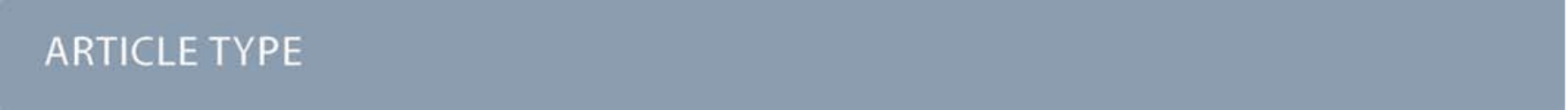}}\par
\vspace{1em}
\sffamily
\begin{tabular}{m{4.5cm} p{13.5cm} }

\includegraphics{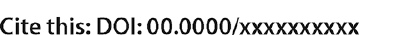} & \noindent\LARGE{\textbf{Fractional Marcus-Hush-Chidsey-Yakopcic current-voltage model for redox-based resistive memory devices}} \\%Article title goes here instead of the text "This is the title"
\vspace{0.3cm} & \vspace{0.3cm} \\

 & \noindent\large{G.~V. Paradezhenko,$^{\ast}$\textit{$^{a}$} D.~V. Prodan,\textit{$^{a}$} A.~A. Pervishko,\textit{$^{a,b}$} D. Yudin,\textit{$^{a,b}$}} and A. Allagui\textit{$^{c,d}$} \\%Author names go here instead of "Full name", etc.

\includegraphics{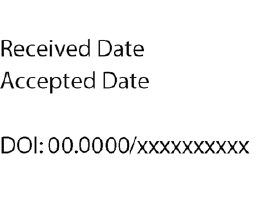} & \noindent\normalsize{We propose a circuit-level model combining the Marcus-Hush-Chidsey electron current equation and the Yakopcic equation for the state variable for describing resistive switching memory devices of the structure metal-ionic conductor-metal. We extend the dynamics of the state variable originally described by a first-order time derivative by introducing a fractional derivative with an arbitrary order between zero and one. We show that the extended model fits with great fidelity the current-voltage characteristic data obtained on a Si electrochemical metallization memory device with Ag-Cu alloy.
}

\end{tabular}

 \end{@twocolumnfalse} \vspace{0.6cm}

  ]
%%%END OF TITLE, AUTHORS AND ABSTRACT%%%

%%%FONT SETUP - please do not change any commands within this section
\renewcommand*\rmdefault{bch}\normalfont\upshape
\rmfamily
\section*{}
\vspace{-1cm}

%%%FOOTNOTES%%%

\footnotetext{\textit{$^{a}$Skolkovo Institute of Science and Technology, Moscow 121205, Russia. E-mail: G.Paradezhenko@skoltech.ru}}
\footnotetext{\textit{$^{b}$ Institute of High Technologies and Advanced Materials, Far Eastern Federal University, Vladivostok 690922, Russia.}}
\footnotetext{\textit{$^{c}$Department of Sustainable and Renewable Energy Engineering, University of Sharjah, Sharjah, P.O. Box 27272, United Arab Emirates.}}
\footnotetext{\textit{$^{d}$Department of Mechanical and Materials Engineering, Florida International University, Miami, FL33174, United States.}}

%%%END OF FOOTNOTES%%%

%%%MAIN TEXT%%%%
\section{Introduction} Substantial research efforts have been dedicated to the development  of electrically-controlled resistive switching in metal-insulator-metal (MIM) devices or memristors, going from new materials discovery \cite{choi2016high, wang20202d, van2018organic, ahn2018carbon, sangwan2020neuromorphic, satapathi2022halide, xu2021design} to modelling and simulation \cite{agudov2020nonstationary, wang2019volatile, zhang2020high}, and design and applications \cite{yang2013memristive, van2018organic, kumar2022dynamical, bao2021discrete}. 
With both memory and logic capabilities combined at the hardware level, in addition to long retention times \cite{yang2013memristive} and high switching rates \cite{torrezan2011sub} at relatively low energy consumption \cite{choi2016high,zhou2016very}, these devices are favorably seen as the next-generation building blocks for nonvolatile memories and neuromorphic computing applications \cite{yang2013memristive, kumar2022dynamical}. 
In a typical memristor, the resistive switching is based on the electrically-stimulated change of cell resistance usually driven by internal ion redistribution, which actually depends not only on the applied excitation but also on the past history of the excitation  \cite{satapathi2022halide}.  
Physical mechanisms associated with these reversible transitions have been attributed to different effects including valence-change\;\cite{wong2012metal}, electrochemical metallization \cite{waser2009redox}, and phase change effects \cite{wong2010phase}. 
They can be either abrupt (binary) or gradual (analogue), and evolve at different timescales, leading to rich and complex device behaviors in this seemingly simple  device structure of just three layers \cite{zidan2018future}. 
Furthermore, with the wide range of diversity in memristors materials and their morphologies, operating mechanisms, and manufacturing technologies there is an urgent need for the development of a general model capable of capturing accurately and effectively their complex nonlinear dynamics. This is crucial not only  for the characterization and comparison between different  memristor devices, but also for the investigation of larger scale memristor-based circuits and  hybrid  hardware architectures, and also to explore  similar behaviors observed for instance in biological synapse systems \cite{ascoli2013memristor}.

While models at different size scales and thus with different degrees of physical details and computational complexity have been developed for memritors, including but not limited to {\it ab initio} \cite{traore2015hfo}, kinetic Monte Carlo, and finite element method models \cite{larentis2012resistive}, in this work we focus on the circuit-level (compact) current-voltage  behavior of memristors.  From this point of view, 
memristors are generally described by the system of coupled equations \cite{chua1971memristor}:
\begin{eqnarray}
  i & = & G(x,v)\cdot v, \label{i-v-eq}\\
  \dot{x} & = & F(x,v), \label{x-eq}
\end{eqnarray}
where $i=i(t)$ is the current through the device, $v=v(t)$ is the applied voltage, and $x=x(t)$ corresponds to a state variable or a group of state variables that quantify the internal dynamics of the device. These are, for example, width of doping region, concentration of vacancies in the gap region, and tunneling barrier width~\cite{agudov2020nonstationary}. State variables can not be observed from external electrical behavior \cite{shang2012analysis}. 
Eq.~\eqref{i-v-eq} follows the $i$-$v$ curve of the resistive device in question with $G(x,v)$ being the generalized conductance, whereas Eq.~\eqref{x-eq} describes the dynamics of its internal state $x$  based on its prehistory \cite{chang2011synaptic}. The actual state of a memristor can only be determined by solving Eqs.~\eqref{i-v-eq} and \eqref{x-eq} self-consistently. 
Memristive systems as featured in terms of Eqs. \eqref{i-v-eq} and \eqref{x-eq} are known to possess a pinched hysteresis loop at the origin in the $i$-$v$ plane in the response to any periodic voltage source~\cite{Chu14}.

Being versatile and modular enough it is the Yakopcic model ~\cite{yakopcic2011memristor, yakopcic2013generalized, yakopcic2019memristor} which is most often used to simulate the nonlinear $i$-$v$ characteristic of wide range of memristors in response to sinusoidal and repetitive sweeping inputs. The model takes into account electron transmission effects, voltage threshold for state variable motion, and nonlinear velocity function for oxygen vacancies or dopant drift, considered to be the most relevant internal state information  \cite{yakopcic2019memristor}. It follows on the steps of Strukov {\it et al.} work~\cite{strukov2008missing}, and describes the  memristor as two resistors in series, one is undoped with high resistance and the other is doped with low resistance, characterized by electron transmission equations so that \cite{yakopcic2019memristor}:
\begin{equation}
  i(t) = h_1(v)x + h_2(v)(1-x).
\label{h12}
\end{equation}
Here, $h_1$ is used to model the behavior in the low-resistance state of the device, and $h_2$ captures its behavior in the high-resistance state.  
The two electron transmission equations are  weighted and mixed by the state variable $x$ which is set to take values between zero and one \cite{chang2011synaptic}.
%, and directly impacts the conductivity of the system. 
In memristive devices, it is the rate of change of the state variable $x$ that is explicitly determined~\eqref{x-eq}, and is given in the   Yakopcic memristor model by the product of the two composite  functions  $g(v)$ and $f(x)$ such that \cite{yakopcic2019memristor}:
\begin{equation}\label{state-eq}
  \dot{x}(t) = g(v) f(x).
\end{equation}
An exponential dependency of the state change to the positive and negative regions of the input voltage $v$ is modelled in terms of
\begin{equation}\label{g-gen-model}
  g(v) =
  \left\lbrace\begin{array}{cc}
        a_p\cdot(1 - e^{u_p-v})\cdot e^v, & v-u_p > 0,\\
        a_n\cdot(e^{u_n+v}-1)\cdot e^{-v}, & v+u_n < 0,\\
        0, &  \textrm{otherwise},
  \end{array}\right.
\end{equation}
including programming voltage thresholds $u_p$ and $u_n$. The magnitude of state change for a voltage potential is defined with $a_p$ and $a_n$.
The second function $f(x)$ is determined by
\begin{equation}\label{f-gen-model-1}
  f(x) =
  \left\lbrace\begin{array}{cc}
        w_p(x,x_p)\cdot e^{-(x - x_{p})}, & x \geq x_p, \\
        1, & x < x_p,
  \end{array}\right. 
\end{equation}
for $v>0$, while for $v<0$, it is defined as
\begin{equation}\label{f-gen-model-2} 
   f(x) =
  \left\lbrace\begin{array}{cc}
        w_n(x,x_n)\cdot e^{x + x_{n} - 1} , & x \leq 1 - x_n, \\
        1, & x > 1 - x_n.
  \end{array}\right.
\end{equation}
Effectively, this function introduces the nonlinear ion motion, as it becomes harder to change the state of the devices when the state variable approaches the boundaries. The degree of this nonlinearity is adjusted by $x_p$ and $x_n$ since the electrode metal used on either side of the dielectric film can react to the dopants differently. In Eq.~\eqref{f-gen-model-1}, $w_p(x,x_p)$ is a windowing function that ensures $f(x)$ equals zero when $x(t) = 1$, and in \eqref{f-gen-model-2}, $w_n(x,x_n)$ keeps $x(t)$ from becoming less than 0 when the current flow is reversed. These two functions can explicitly be written as $w_p(x,x_p)=1+(x_p - x)/(1 - x_p)$ and $w_n(x,x_n)=x/(1 - x_n)$.

Clearly, in~\eqref{h12}, the functions $h_{1}$ and $h_{2}$ depend on the structure and type of the memristor under study. Several types of resistive switching memory devices can be classified as nanoionic-based electrochemical systems, wherein an ion conductor in the form of electron insulator layer is placed between two electrodes~\cite{valov2013nanobatteries, valov2011electrochemical}.   For the case of cation-migration-based electrochemical metallization memory cells, \ce{Ag} or \ce{Cu} are  typically used as active electrodes, \ce{Pt} or \ce{W} as counter electrodes, and a variety of oxides or chalcogenides thin films as solid electrolytes. When a  positive voltage is applied, the active electrode material is oxidized at the electrode-electrolyte interface leading to the release of metallic ions in the adjacent electrolyte, followed  by drift and diffusion of these  ions across the electrolyte, and then  their deposition in filamentary-like metal structures at the counter electrode surface. Short-circuit occurs when the filament has grown sufficiently far to make an electronic  contact with the opposite electrode, which defines the low-resistance state  of the cell. When a negative voltage is applied, the cell returns back, in principle reversibly, to the high-resistance state~\cite{valov2013nanobatteries}.  
Anion-migration-based valence change cells, on the other hand, are formed by placing a metal oxide  between for example \ce{Pt} or \ce{TiN} electrodes and another oxygen-affine, lower work function electrode. The low-resistance and high-resistance states are defined based  on the electrochemical formation of oxygen-deficient, mixed ionic-electronic conducting filaments, and the nanoionic modification of the potential barrier between the tip of the filament and the electrode it faces  \cite{valov2013nanobatteries}. For  these types of redox-based resistive memory cells, it is more appropriate  to consider electron transfer theory associated with the kinetics of redox reactions to better describe their $i$-$v$ characteristics. Furthermore, because the formation and rupture of the metallic filaments follow random paths, the possibility of charge trapping from one operation sequence to another, charge leakage, the dynamics of an internal state variable  associated with these cells cannot be defined solely based on its immediate past, in other words via integer-order derivative as in~\eqref{x-eq}. Taking into account the integral past is believed to be more representative for a proper mathematical description of the complexity and dissipative nature of these cells.   

Motivated by these observations, we herein propose a circuit-level model for redox-based resistive memory devices, where the current equation~\eqref{i-v-eq} is taken from the Marcus-Hush-Chidsey (MHC) theory~\cite{Chi91,Mar96,ZSB14} of heterogeneous electron transfer, while the state variable equation~\eqref{x-eq} is taken from the Yakopcic generalized memristive model~\cite{yakopcic2011memristor}. We  consider the dynamics of the state variable with respect to time to be of fractional, non-integer, order. Mathematically, this adds an extra degree of freedom to the model that can be generically correlated to the non-perfect reversibility of the device when looking at it from one cycle to another. We fit the extended model to the experimental data obtained on a \ce{Si} memristor with Ag-Cu alloy, shown in Fig.~\ref{fig:device}, as reported in~\cite{KAA22}, and make a direct comparison with the superstatistics approach developed therein. A close inspection of numerical results unambiguously reveals that switching to the fractional derivative allows one to significantly improve the agreement between the theory and experimental data.

\section{Memristor model} 

The generalized $i$-$v$ relationship for the proposed memristor model is specified by Eq.~\eqref{i-v-eq} with
\begin{equation}\label{h-gen-model}
  h_{j}(v)=\gamma_j\cdot h(\delta_j\cdot v),
  \qquad 
  j = 1,2,
\end{equation}
where $\delta_1, \delta_2, \gamma_1, \gamma_2 >0$ are fitting parameters. As a rule, these parameters are material-specific and temperature-dependent, so that $\delta_1$ and $\delta_2$ can be viewed as magnitudes of the current conductivities, while $\gamma_1$ and $\gamma_2$ control the curvatures in the $i$-$v$ curve relative to the applied voltage $v$. 
The function
\begin{equation}\label{h-def}
  h(v) = h_{+}(v) - h_{-}(v),
\end{equation}
is based on the MHC model for electron transfer described by the Gauss-Fermi integral~\cite{ZSB14},
\begin{equation}\label{MHC-func}
  h_{\pm}(v) = \beta \int_{-\infty}^{\infty} \exp\left\lbrace -\frac{(z - \lambda \pm v )^2}{4 \lambda}  \right\rbrace \frac{dz}{1 + e^z}.
\end{equation}
Here, the $\pm$ signs refer to the oxidative and reductive transition rate functions, $\lambda$ is the dimensionless reorganization energy scaled to $k_{\rm B} T$, while the integral over the dimensionless variable $z$ accounts for the Fermi statistics of electron energies, distributed around the electrode potential. The prefactor $\beta$ specifies the electronic coupling strength and the electronic density of states of the electrode. In Eq.~\eqref{MHC-func}, $\lambda$ and $\beta$ are assumed to be fitting parameters, knowing that $\beta$ is usually expressed as an exponential term itself that depends on the distance between the donor and acceptor of electrons. This, however, does not affect the generality of the proposed model. Finally, $v$ in Eqs.\;\eqref{i-v-eq}--\eqref{MHC-func} is actually the electrochemical overpotential defined as the difference between the equilibrium Nernst-potential  of the metal and the actual electrode  potential defined by the external power supply. We will consider  the equilibrium potential   to be negligible, so that the electrochemical potential is equal to the   applied voltage on the device.

For the dynamics of the state variable $x(t)$, we introduce a fractional time derivative in \eqref{state-eq} as follows,
\begin{equation}\label{state-eq-fracder}
  D_t^{\alpha} x(t) = g(v) f(x),
  \qquad
  x(0) = x_0,
\end{equation}
where $D_t^{\alpha}$ is the fractional derivative operator of order $\alpha>0$ in the sense of Caputo,
\begin{equation}\label{Caputo}
  D_t^{\alpha}x(t) \equiv \frac{1}{\Gamma(m-\alpha)} \int_0^t \frac{x^{(m)}(\tau) d\tau}{(t-\tau)^{\alpha+1-m}},
\end{equation}
where $m = \lceil\alpha\rceil$, while $\Gamma(z)=\int_0^\infty t^{z-1}e^{-t}dt$ in the denominator stands for the Gamma function, and $x^{(m)}(\tau)$ is the $m$-th order derivative. In our present study, the order of the  derivative is assumed to be $0<\alpha<1$. 

Non-integer $\alpha<1$ in~\eqref{Caputo} for the state variable equation~\eqref{state-eq-fracder} indicates that its dynamics does not evolve without prior knowledge of all past information of its state. Or in other words, the right-hand side of~\eqref{Caputo} contains information about all the previous states of the physical system, representing the so-called memory trace~\cite{teka2014neuronal}. The memory trace term increases with $\alpha$ decreasing away from one, and when we are close to the actual instant $t$ at which the variable is evaluated. At the limiting case of $\alpha=1$, corresponding to first-order integer derivative, the memory trace part vanishes, and does not have any effect on the dynamics of the state variable.

Thus, the parameters of the proposed memristor model can be summarized into a single vector 
\begin{equation}\label{parameters}
  \bm{p} = (\alpha, x_p, x_n, a_p, a_n, u_p, u_n, \beta, \lambda, \gamma_1, \gamma_2, \delta_1, \delta_2).
\end{equation}

\section{Methods}

\subsection{Evaluation of the MHC integral}
The Gauss-Fermi integrals like \eqref{MHC-func} can be evaluated numerically using the Gauss-Hermite quadrature (see, e.g., \cite{PTV17}). In practice, 
\begin{equation}\label{quadrature}
	\int_{-\infty}^{\infty} e^{-z^2} q(z)\, d z \approx \sum_{k=1}^n c_k q(z_k),
\end{equation}
where $n$ corresponds to the amount of sample points, $q(z)$ is an arbitrary function, while $z_k$ are the roots of the Chebyshev-Hermite polynomial $H_n(z_k)=0$ with $k=1,\ldots,n$. For a given $n\geq2$ the Chebyshev-Hermite polynomial $H_n(z)$ can be identified from recurrence relations
\begin{equation}
    H_{n+1}(z)=2zH_n(z)-2nH_{n-1}(z),
\end{equation}
provided $H_0(z)=1$ and $H_1(z)=2z$. In~\eqref{quadrature}, the coefficients $c_k$ are given by
\begin{equation}
    c_k = \frac{2^{n-1}}{n^2}\frac{n!}{[H_{n-1}'(z_k)]^2}\sqrt{\pi}.
\end{equation}
Thus, one can rewrite the MHC integrals~\eqref{MHC-func} in the form
\begin{equation}
    h_{\pm}(v) =  2\beta\sqrt{\lambda} \sum_{k=1}^n\frac{c_k}{1+\exp\{2x_k\sqrt{\lambda}+\lambda\pm v\}}.
\end{equation}
In our numerical simulations, the order of quadrature $n=25$, which is deemed more than sufficient  for our purpose. 

\subsection{Solution to the fractional differential equation}

The nonlinear fractional differential equation \eqref{state-eq-fracder}
is solved numerically using the Adams-type predictor-corrector method \cite{DFF02}.
For  nonlinear fractional differential equations of the form
\begin{equation}\label{fracder}
  D_t^{\alpha} x(t) = F(t,x),
  \quad
  x^{(k)}_{}(0) = x^{(k)}_0, 
\end{equation}
where $k=0,1,\ldots,m-1$ and $m = \lceil\alpha\rceil$,
this method can be described as follows. The approach is based on the fact that the initial value problem is 
equivalent to the Volterra integral equation
\begin{equation}\label{Volterra}
  x(t) = \sum_{k=0}^{m-1}\frac{t^k x^{(k)}_0}{k!} + \frac{1}{\Gamma(\alpha)} \int_0^t \frac{F(\tau, x(\tau))}{(t - \tau)^{1-\alpha}}\, d\tau.
\end{equation}
We assume that the choice of $F(t,x)$ guarantees the existence of a unique solution in a certain interval $0\leq t\leq T$. We divide this interval into $N$ equal pieces as specified by a uniform grid at the points $t_n=hn$, $n=0,1,\ldots,N$ and $h=T/N$. The basic idea is that using pre-calculated approximations $x_h(t_j) \approx x(t_j)$, $j=0,1,\ldots,n$,
we get the next time step approximation $x_h(t_{n+1})$ by means of Eq.~\eqref{Volterra}.
 
Replacing the integral on the right-hand side of Eq.~\eqref{Volterra} by the product rectangle rule, we obtain
\begin{equation}\label{Volterra-approx-pred}
  \int_0^{t_{n+1}} \frac{F(\tau,x(\tau))}{(t_{n+1} - \tau)^{1-\alpha}}\, d\tau \approx \frac{h^\alpha}{\alpha} \sum_{j=0}^n F_j\cdot b_{n-j},
\end{equation}
where $F_j = F(t_j,x_h(t_j))$ and $b_{k} = (k+1)^\alpha - k^\alpha$, provided that $0\leq k \leq n$. The predicted value $x^P(t_{n+1})$ is determined by the fractional Adams-Bashforth method,
\begin{equation}\label{predictor}
  x_h^P(t_{n+1}) = \sum_{k=0}^{m-1}\frac{t_{n+1}^kx^{(k)}_0}{k!} 
  + \frac{h^\alpha}{\Gamma(\alpha+1)} \sum_{j=0}^nF_j\cdot b_{n-j}.
\end{equation}
To obtain a formula for the corrector, one uses the product trapezoidal quadrature formula to replace the
integral in Eq.~\eqref{Volterra}, where nodes $t_j$ are taken with respect to the weight function $(t_{n+1} - \tau)^{\alpha-1}$.
Using standard techniques from quadrature theory, we can write the integral on the right-hand side of Eq.~\eqref{Volterra} as
\begin{equation}\label{Volterra-approx-corr}
  \int_0^{t_{n+1}}\frac{F(\tau,x(\tau))}{(t_{n+1} - \tau)^{1-\alpha}}\, d\tau \approx \frac{h^\alpha}{\alpha(\alpha+1)}\sum_{j=0}^{n+1}F_j\cdot a_{n-j},
\end{equation}
where $a_{-1}=1$ and $a_n=n^{\alpha+1}-(n-\alpha)(n+1)^\alpha$, while $a_k=(k+2)^{\alpha+1}-2(k+1)^{\alpha+1}+k^{\alpha+1}$ for $k=1,\ldots,n-1$. We thus come to the corrector approximation, which can be thought of as the fractional variant of the one-step Adams-Moulton method, 
\begin{eqnarray}
  x_h(t_{n+1}) &=& \sum_{k=0}^{m-1}\frac{t_{n+1}^kx^{(k)}_0}{k!} 
  + \frac{h^\alpha}{\Gamma(\alpha+2)} F(t_{n+1}, x_h^P(t_{n+1})) \nonumber \\
  &+& \frac{h^\alpha}{\Gamma(\alpha+2)} \sum_{j=0}^{n}F_j\cdot a_{n-j}.
  \label{corrector}
\end{eqnarray}
The numerical error of this method is shown to behave as
\begin{equation}\label{error}
  \max_{n=0,1,\ldots,N} \bigl| x(t_n) - x_h(t_n)  \bigr| = O(h^p),
\end{equation}
with $p = \min\lbrace 2, 1+\alpha\rbrace$. In practice, we first calculate and store the coefficients given by $\{b_k\}$ and $\{a_k\}$ of Eqs.~\eqref{Volterra-approx-pred} and \eqref{Volterra-approx-corr} as arrays. 
After that, on each time step we calculate the predictor  \eqref{predictor} and then use it to calculate the corrector \eqref{corrector}. 
To speed up the calculations, we apply the Fast Fourier Transform  algorithm to compute the convolutions on the right-hand sides of Eqs.~\eqref{predictor} and \eqref{corrector}.

\subsection{Fitting method}

Suppose the $i$-$v$ curve is yielded by $N$ measurements 
$\{(t_k, v_k, i_k)\}_{k=1}^N$, where $v_k = v(t_k)$ and $i_k = i(t_k)$. 
To fit the model as specified by Eqs.~\eqref{i-v-eq} and \eqref{state-eq-fracder} to this data, we search for the set of fitting parameters \eqref{parameters} using the least squares method. This is done by applying the Trust Region Reflective algorithm~\cite{BCL99} to minimize the cost function,
\begin{equation}\label{ls-cost}
  \bm{p}^\ast
  = \mathrm{arg}\,\min_{\bm{p}}\sum_{k=1}^N \left[ i_k - i^\mathrm{mod}(v_k, t_k, x(t_k), \bm{p}) \right]^2 ,
\end{equation}
where $i^\mathrm{mod}(v,t,x,\bm{p})$ is the model current specified by the right-hand side of~\eqref{i-v-eq}. The parameters are non-negative, and the fractional derivative order is bounded, $0<\alpha\leq1$. Since the current~\eqref{i-v-eq} depends on the state variable $x(t)$, for each $\bm{p}$ we self-consistently solve either the ordinary~\eqref{state-eq} or fractional~\eqref{state-eq-fracder} differential equation with respect to the state variable in $0\leq t\leq T$. As long as the evolution of $x(t)$ is described in terms of the ordinary differential equation, we keep the parameter $\alpha=1$ excluded from the fitting parameters vector~\eqref{parameters}. Eq.~\eqref{state-eq} is solved numerically using the Runge-Kutta-Fehlberg method, while Eq.~\eqref{state-eq-fracder} is addressed by a means of the Adams-type predictor-corrector method on condition that $x(0)=0$. Once the state variable $x(t)$ is calculated, we interpolate it at time steps $t_k$ and evaluate the current  $i^\mathrm{mod}(v,t,x,\bm{p})$ for specific points $\{(v_k, t_k, x(t_k))\}_{k=1}^N$. The fitted model is then evaluated and compared to the experimental data using the Normalized Root-Mean-Square Error (NRMSE),
\begin{equation}\label{NRMSE}
  \mathrm{NRMSE} = \frac{\frac{1}{\sqrt{N}} \sqrt{\sum_{k=1}^N (i_k^{} - i^\mathrm{mod}_k} )^2}{\frac1N \sum_{k=1}^N {i_k}},
\end{equation}
where $i_k^\mathrm{mod} = i^\mathrm{mod}(v_k, t_k, x(t_k),\bm{p})$ is the evaluated model current.

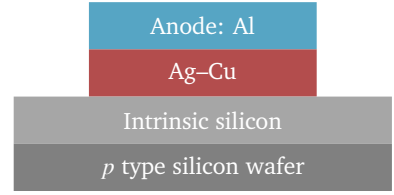
\begin{figure}
	\centering
	\begin{tikzpicture}
		\node[draw=gray!50!cyan, fill=gray!50!cyan, thick, minimum height=0.6cm, minimum width=3cm](anode) at (0,0) {\color{white}{Anode: Al}};
		\node[draw=gray!60!red, fill=gray!60!red, thick, minimum height=0.6cm, minimum width=3cm, below = 0cm of anode](dopant) {\color{white}{Ag--Cu}};
		\node[draw=black!35, fill=black!35, thick, minimum height=0.6cm, minimum width=5cm, below = 0cm of dopant](carrier){\color{white}{Intrinsic silicon}};
		\node[draw=black!50, fill=black!50, thick, minimum height=0.6cm, minimum width=5cm, below= 0cm of carrier](Cathode: ){\color{white}{$p$ type silicon wafer}};
	\end{tikzpicture}
\caption{A schematic showing the layer-by-layer structure of the memristor under test.}
\label{fig:device}
\end{figure}

\begin{table}[ht!]
\small
\caption{Fitting results for the MHC-Yakopcic model with ordinary and fractional differential equation for the state variable.\label{tab:params}}
\begin{tabular*}{0.48\textwidth}{l@{\extracolsep{\fill}}ccc}
\hline
Parameters & Integer order & Fractional order \\
\hline
\hline
$\alpha$ & 1.000 & 0.677 \\
$x_p$ &  0.587 & 0.577 \\
$x_n$ & 0.0 & 0.989 \\
$a_p$ & 0.068 & 0.064 \\
$a_n$ & 0.093 & 0.006\\
$u_p$ & 2.373 & 4.883 \\
$u_n$ & 0.000 & 0.000 \\
$\beta$ & 33.37 & 1.377 \\
$\lambda$ & 28.27 & 17.40 \\
$\gamma_1$ & 30.17 & 1.743 \\
$\gamma_2$ & 3.663 & 2.567 \\
$\delta_1$ & 1.072 & 4.509 \\
$\delta_2$ & 1.597 & 2.315 \\ 
\hline
NRMSE & 0.781 & 0.776 \\
\hline
\end{tabular*}
\end{table}

\begin{figure*}[t!]
\centering
\includegraphics[width=1.0\linewidth]{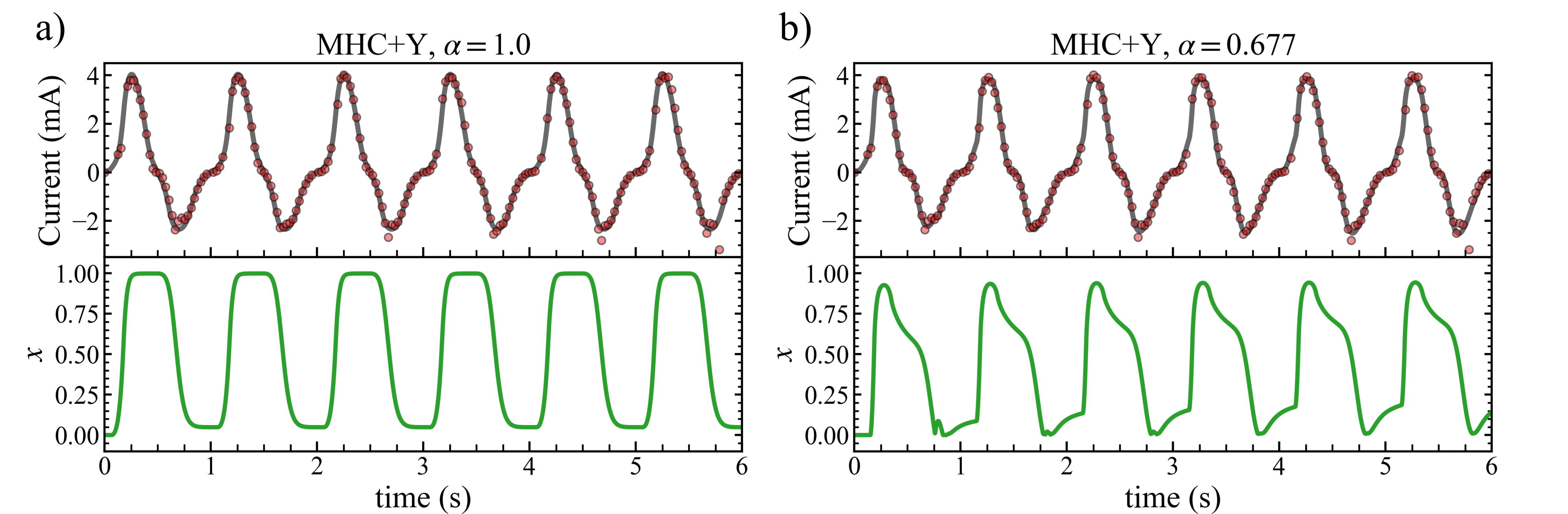}
\caption{Current $i(t)$ and state variable $x(t)$ as calculated by the MHC-Yakopcic model fitted to the data of the memristor under study, provided the dynamics of the state variable is described in terms of a) the ordinary differential equation~\eqref{state-eq} and b) the fractional differential equation~\eqref{state-eq-fracder}. The experimental data for the current is depicted by the red points.}
\label{fig:ix-calc}
\end{figure*} 

\section{Results and discussion}

We fitted the memristor specified by Eqs.~\eqref{i-v-eq} and \eqref{state-eq-fracder} combining the MHC-based state-controlled current-voltage relationship and the fractional Yakopcic state variable model to the $i$-$v$ characteristic data of the electrochemical metallization memory device taken from \cite{KAA22}. The device is a Si memristor with \ce{Ag}–\ce{Cu} alloying conducting channels that was fabricated following the method of Yeon {\it et al.}~\cite{yeon2020alloying}. A schematic of the fabricated memristor can be seen in Fig.~\ref{fig:device}. For the $i$-$v$ measurements that were carried out on a BioLogic VSP-300  workstation, six successive  sinusoidal voltage waveforms were applied across the two terminals of the device such that 
\begin{equation}
  v(t) = u_0 \sin (2\pi f t),
\end{equation}
with $u_0=6$~V and $f=1$~Hz in the time course $0\leq t\leq 6$~s. The fitting is then performed on this time interval to the whole six cycles of switching~\cite{Prodan2023}. Note that the fitting method is implemented on C++ using the FFTW library~\cite{FFTW} for Fast Fourier Transform and least-squares routine from the GSL~\cite{GSL}. The implemented framework allows a model with a single set of parameters to be fitted to multiple datasets at once, so that the main procedure returns a complete array of all fitting parameters and costs for each performed run. Noteworthy, our findings suggest the obtained value of $\alpha$ is robust to the variance of the input data.

The numerical simulations of the models with the ordinary and fractional derivatives fitted to the experimental data are presented in Fig.~\ref{fig:ix-calc}. Here, we show the numerical solution of the canonical memristor system \eqref{i-v-eq} and \eqref{x-eq} for the MHC-Yakopcic model in terms of $i(t)$ and $x(t)$, where the calculated current is compared to the measured one.
As mentioned above, we considered the equilibrium Nernst-potential  of the electrode to be zero, so that the electrochemical potential in Eqs.~\eqref{i-v-eq} and \eqref{x-eq} is equal to the actual applied voltage on the device. The corresponding fitting parameters are provided in Table~\ref{tab:params}.
As one can see, the MHC-Yakopcic model fits very well to the experimental data with NRMSE~$=0.776$ for the fractional order $\alpha=0.677$. Remarkably enough, NRMSE~$=0.781$ when the state variable evolves according to the ordinary differential equation, {\it i.e.}, with $\alpha=1$. It is worth mentioning that averaging over cycles is typically done in literature to show a characteristic curve of the device under study. Meanwhile, the device history, or memory, effect is blurred when we try to fit averaged data. Instead, in our analysis, we fit the proposed model to the whole six cycles of switching as shown in Fig.~\ref{fig:ix-calc}. As the result, our model with $\alpha<1$ shows improvement in terms of the NRMSE against the same model with fixed $\alpha=1$. This improvement demonstrates that the fractional dynamics could be valuable enough for describing memristive switching, provided ageing and effects of degradation.

In comparison, we evaluated the $q$-deformed memristor model recently reported in~\cite{KAA22} using the provided parameters and got NRMSE~$=0.827$. Note that this value is almost twice larger than the reported one in Ref.~\cite{KAA22} since the fit was performed to the data averaged over six cycles of switching. The $q$-deformed model was derived by taking into account gamma-distributed local spatial inhomogeneities in the device structure. This provided  a noticeable improvement in the fitting of the $i$-$v$ response of the same device under study here when compared to the currently used existing  model ({\it i.e.},  the Yakopcic model with MIM and  Schottky electron transmission equations)~\cite{KAA22}. In practice, it was shown that by introducing a single additional parameter that is related to the $q$-deformed exponential, one can achieve better agreement with the experimental data for the device under study. This $q$-parameter being of pure mathematical nature can be however associated with the fractional order $\alpha$~\cite{Herrmann2010}. In the meantime, as mentioned above, the kinetics of  electrode reactions in redox-based electrochemical metallization memory cells should be rather described by electron transfer theory. That is, the MHC+Yakopcic model provides about 8\% improvement in comparison to the $q$-deformed model.

\begin{figure}[!t]
\centering
\includegraphics[width=0.85\linewidth]{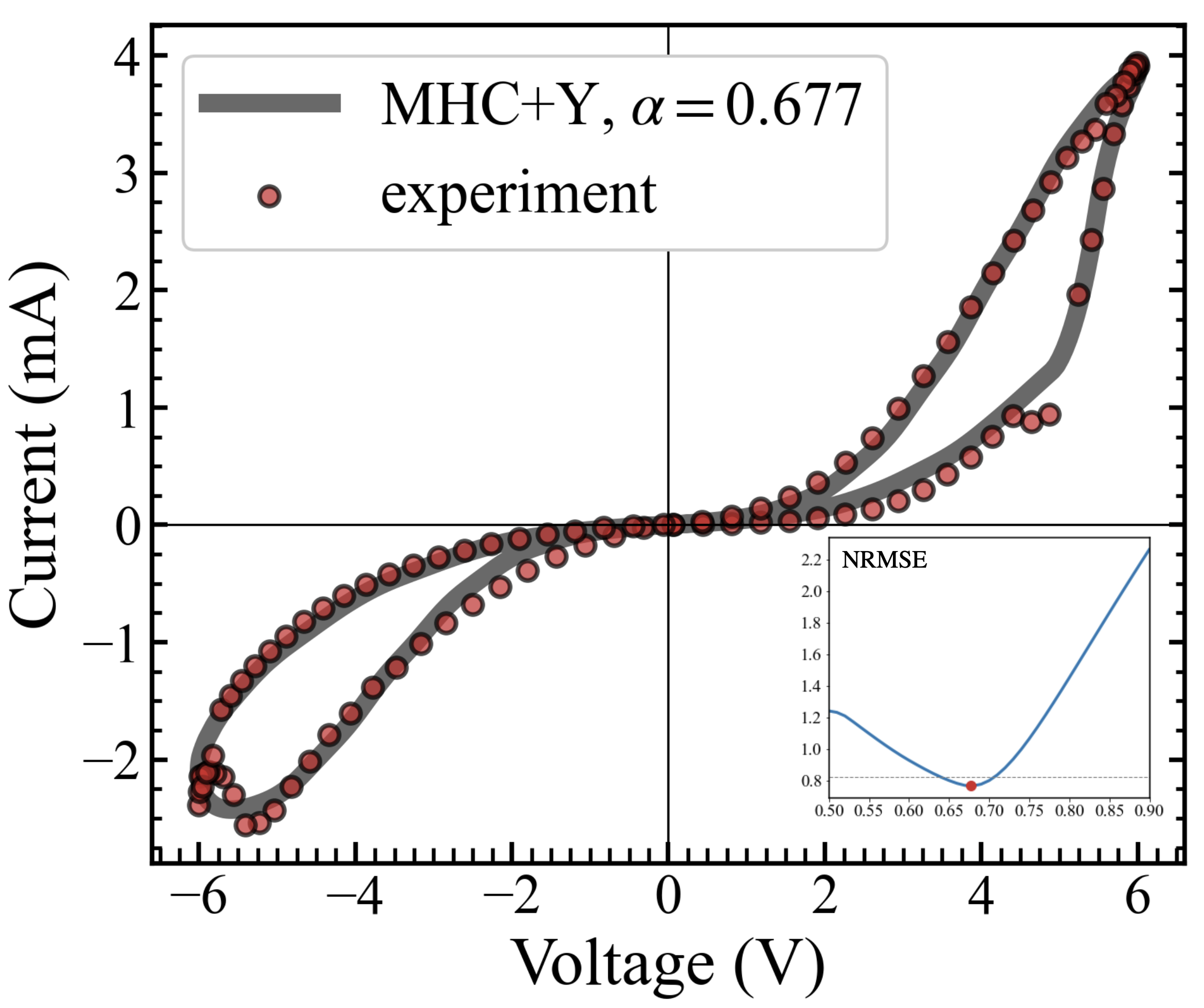}
\caption{Current-voltage characteristic calculated by the fitted MHC-Yakopcic model with fractional order dynamics in comparison with the experimental data. The data is averaged over six cycles of switching for both experimental and model $i$-$v$ characteristic. The NRMSE, shown in inset, reaches its minimum at $\alpha=0.677$ as marked by the red dot. Remarkably, the NRMSE exceeds 0.827 obtained as a result of fitting the $q$-deformed model to the same data in the range of $\alpha<0.64$ and $\alpha>0.71$.}
\label{fig:iv-calc}
\end{figure} 

Here, because $\alpha \neq 1$ we may speak of an intrinsic  memory embedded in our redox-based resistive memory device. Fractional dynamics is in fact very often observed in electrochemical devices and complex systems~\cite{tarasov2011fractional, allagui2022extended, allagui2022power, hernandez2020exploring, zhang2019modeling, luchko2010fractional,  metzler2000random}. Interestingly enough, the saturation time of a typical memristor that is needed to bring it from the low resistance state to the high resistance state under the applied voltage is sensitive to the fractional order $\alpha$~\cite{Wang2020}. Thus, the latter, in principle, can be identified from knowing the saturation time.

\section{Conclusions}

In this work we proposed a compact and accurate model for describing the electrical behavior  of redox-based resistive memory devices in which (i) the state-controlled current-voltage equation is based on the MHC theory for electron transfer, and (ii) the dynamics of the state variable is assumed to follow fractional time  derivatives of order $\alpha$ ($0<\alpha<1$), with the latter adding a {\it non-Markovian} or {\it memory trace} term to the modeled dynamics. For the numerical solution to the MHC integral we used the Gauss-Hermite quadrature method and for the fractional differential equation of the state variable we used an Adams-type predictor-corrector technique. Goodness of fit to the experimental data is evaluated in terms of NRMSE, and indicates advanced capabilities of the proposed model when compared to recently reported ones. The developed numerical routine allows one to uniquely determine the value of $\alpha$. It should be stressed that the NRMSE is used herein for evaluation to produce a normalized, dataset-independent quality metric for the tested model. This allows for a more standardized way of comparing models between this work and other studies. The obtained results, in connection to the electrochemical nature of the device under test, point out to necessity to take into consideration fractional dynamics, that could be of importance provided ageing and general degradation of the device, when describing the $i$-$v$ characteristics of redox-based resistive memory cells. We showed that the proposed model with fractional order $\alpha<1$ dynamics provides 1\% improvement in terms of the NRMSE in comparison to the model with fixed $\alpha=1$. It is worth mentioning that the $q$-number as introduced previously in the spirit of superstatistics approach~\cite{KAA22} is can be linked to the fractional order $\alpha$~\cite{Herrmann2010}.

\section*{Conflicts of interest}
There are no conflicts to declare.

\section*{Acknowledgements}
A.A.P. and D.Y. acknowledge the support from the Russian Ministry of Science and Higher Education Project No. 075-15-2021-607.
%%%END OF MAIN TEXT%%%


\balance


\providecommand*{\mcitethebibliography}{\thebibliography}
\csname @ifundefined\endcsname{endmcitethebibliography}
{\let\endmcitethebibliography\endthebibliography}{}

\end{document}